\documentclass[aip,jcp,reprint]{revtex4-1}

\usepackage{dcolumn}       
\usepackage{bm}            
\usepackage{amsmath}       
\usepackage{graphicx}      
\usepackage{natbib}		   
\usepackage{amssymb}       
\usepackage{latexsym}      
\usepackage{multirow}      
\usepackage{array}         
\usepackage{wrapfig}       

\newcommand{\bs}[1] {  \boldsymbol{#1}           }
\newcommand{\ff}[1] {  \mbox{\footnotesize{#1}}  }
\newcommand{\Ang}   {  \mbox{\normalfont\AA}     }

\begin{document}
\title{Polar nanoregions in water - a study of the dielectric properties of TIP4P/2005, TIP4P/2005f and TTM3F}

\author{D. C. Elton}
\affiliation{Department of Physics and Astronomy,  Stony Brook University, Stony Brook, New York 11794-3800, USA}

\author{M.-V. Fern\'{a}ndez-Serra} 
\affiliation{Department of Physics and Astronomy,  Stony Brook University, Stony Brook, New York 11794-3800, USA}
\affiliation{Institute for Advanced Computational Sciences, Stony Brook University, Stony Brook, New York 11794-3800, USA}

\date{\today}
\begin{abstract}
We present a critical comparison of the dielectric properties of three models of water - TIP4P/2005, TIP4P/2005f and TTM3F. Dipole spatial correlation is measured using the distance dependent Kirkwood function along with one dimensional and two dimensional dipole correlation functions. We find that the introduction of flexibility alone does not significantly affect dipole correlation and only affects $\varepsilon(\omega)$ at high frequencies. By contrast the introduction of polarizability increases dipole correlation and yields a more accurate $\varepsilon(\omega)$. Additionally the introduction of polarizability creates temperature dependence in the dipole moment even at fixed density, yielding a more accurate value for $d \varepsilon / d T$ compared to non-polarizable models. To better understand the physical origin of the dielectric properties of water we make analogies to the physics of polar nanoregions in relaxor ferroelectric materials. We show that $\varepsilon(\omega,T)$ and $\tau_D(T)$ for water have striking similarities with relaxor ferroelectrics, a class of materials characterized by large frequency dispersion in $\varepsilon(\omega,T)$, Vogel-Fulcher-Tamann behaviour in $\tau_D(T)$, and the existence of polar nanoregions. 
\end{abstract}

\maketitle

\section{Introduction}
Water's dielectric properties are central to understanding water's role as a solvent and are important in areas such as climate science, remote sensing and microwave engineering. The great practical importance of water's dielectric properties has led to their measurement to high accuracy at a large gamut of state points.\cite{fernandez:1125,ellison:1,Ellison:0167}

A central question we seek to answer is what the relative effects of water model geometry, flexibility and polarization are on the dielectric constant. The usefulness of adding flexibility to water models has been investigated before with mixed results,\cite{Tironi:19,Teleman:193,wu:024503,Dinur:5669,Ferguson:1096} and many polarizable models have likewise been created and investigated.\cite{Sprik:7556,Zhu:6211,lamoureux:5185,yu:221,kumar:014309,Ren:5933,vanMaaren:2618,li:154509,Wang:9956,
Troster:9486,Stern:2001} Critical comparisons of rigid vs. flexible and/or polarizable models have been done before with a focus on reproducing the density anomaly,\cite{mahoney:10758} IR spectra,\cite{Hasegawa:5545} water clusters,\cite{yu:9549} and H-bond dynamics.\cite{Xu:2054}  In this paper we examine the importance of both flexibility and polarizability on the dielectric properties of water. We do this by comparing three models with similar geometries - the rigid and flexible versions of TIP4P/2005 and TTM3F, which is flexible and polarizable.

In the process of comparing these three models we compare the nature and degree of dipolar correlation in detail and investigate how this correlation contributes to the dielectric properties. To better understand the dielectric properties as a whole we ask if water can be understood as a relaxor ferroelectric. Relaxor ferroelectrics are highly polarizable materials characterized by broad temperature dispersion in $\varepsilon(\omega, T)$ and the presence of polar nanoregions.\cite{Samara:R367,Kleemann:275}

\subsection{Dipolar correlations in water}

Water is exceptional in its ability to form highly ordered phases under certain conditions. Most strikingly, Ice XI is a proton-ordered ferroelectric phase which forms when Ice Ih is cooled below 72 K. Local ferroelectric ordering is preserved when Ice XI is transformed into Ice Ih, leading to easier reformation of Ice XI upon recooling.\cite{Arakawa:111} Water confined in carbon nanotubes or membrane channels is believed to exhibit ferroelectric order.\cite{Nakamura1064,Luo:2607,Kofinger:36}
The presence of an interface is known to influence the structure of water and degree of dipolar correlation up to several nanometers into the bulk.\cite{Zhang:2477,lee:4448,zhu:1403,Kanth:021201} 
In biophysics, some proteins have ``ferroelectric" hydration shells with thicknesses of 3-5 water diameters,\cite{LeBard:9246} and antifreeze proteins are believed to influence water structure up to a nanometer into the bulk.\cite{Meister:31122012} 

In bulk water the degree of dipolar correlation is well quantified by the Kirkwood factor $G_K$. Assuming conducting boundary conditions, the dielectric constant can be calculated in a computer simulation using the following linear response relation:
\begin{equation}\label{eqnused}
    \varepsilon(0) -  \varepsilon_\infty  = \frac{1}{3 k_B \epsilon_0 T V} \left( \langle {\bf M}^2 \rangle - \langle {\bf M} \rangle^2  \right)
\end{equation}
Here ${\bf M} = \sum_i^N {\bf \mu}_i$ is the total dipole moment of the simulation box. $\varepsilon_\infty = 1$ for a rigid model and can be well estimated using the Clausius-Mossotti relation for flexible and polarizable models.\cite{Neumann:polar,lamoureux:5185} To see the dependence of $\varepsilon(0)$ on dipolar correlation it is useful to recast equation \ref{eqnused} as:
\begin{equation}\label{eqnused2}
        \varepsilon(0) - \varepsilon_\infty  = \frac{N \mu^2}{3 k_B \epsilon_0 T V} G_K \\
\end{equation}
Here $G_K$ is the finite system g-factor. If we assume $ \langle {\bf M} \rangle^2 \rightarrow 0$ then
\begin{equation}
            \varepsilon(0) - \varepsilon_\infty  = \frac{N \mu^2}{3 k_B \epsilon_0 T V} (1 + N \langle \cos (\theta) \rangle)
\end{equation}
$\langle \cos (\theta) \rangle$ is the average cosine of the angle between dipoles. It is important to distinguish between the finite system Kirkwood factor $G_K$ and infinite system Kirkwood factor $g_K$.\cite{Neumann:97} $g_K$ was defined by Kirkwood as:\cite{K39}
\begin{equation}\label{KirkwoodEqn}
    \frac{(\varepsilon(0) - 1)(2\varepsilon (0)  + 1) }{3 \varepsilon (0)} = \frac{N \mu^2}{3 k_B \epsilon_0 T V} g_K
\end{equation}
This equation is the exact equation for rigid dipoles in an infinite medium with no boundary at infinity.\cite{hansen2006theory,K39} The relation between $g_K$ and $G_K$ varies considerably depending on the boundary conditions and method employed for treating the long range interactions.\cite{Neumann:97} For Ewald summation with conducting boundary conditions:\cite{Neumann:97}
\begin{equation}\label{Gkgkrelation}
    g_K = \frac{2\varepsilon (0) + 1}{ 3 \varepsilon (0) } G_K
\end{equation}
Note that equations \ref{KirkwoodEqn} and \ref{Gkgkrelation} are only strictly correct for rigid dipoles ($\varepsilon_\infty = 1$), but we found that the correction to \ref{KirkwoodEqn} from polarization contributes negligibly to $g_K$ (about 1.5 \%).\cite{Neumann:polar} Looking at equation \ref{eqnused2} we see that if the dipoles are uncorrelated ($\langle \cos (\theta) \rangle = 0$) then $G_K = 1$ and $\varepsilon (0)$ would equal 30 for water at 298 K (assuming a dipole of $2.95$ D). The actual value is 78.4, indicating that dipolar correlations increase $\varepsilon(0)$ by a factor of $G_K = 2.6$.

In bulk water it is well known that the tetrahedral hydrogen bond network increases dipolar correlation.\cite{K39} If we assume a four-site tetrahedral bonding model with bonding probability $P$ and ignore all H-bond loops, then the contribution of the $i$th H-bonded shell to $G_K$ is given by:\cite{suresh:9727} 
\begin{equation}\label{HbondShells}
    4P^i \cos^2 (\theta_{\ff{HOH}}/2) \cos (\pi - \theta_{\ff{HOH}})^{i-1}
\end{equation}
Assuming $\theta_{\ff{HOH}} = 109\,^{\circ}$ and $P = .875 $ then this yields $G_K = 2.65$ with contributions of $G_K - 1 = 1.18 + .34 + .09 + .03 + \cdots $. 

The importance of the H-bond network is confirmed in computer simulations which show a strong correlation between hydrogen bond density and dielectric constant.\cite{sprik:6762,Yoshii:195} The importance of the H-bond network can also be inferred from the observation that dissolved solutes decrease $\varepsilon(0)$.  Remarkably, the decrease in $\varepsilon(0)$ with solute concentration is largely independent of the type of solute,\cite{Kaatze:1049} suggesting that the depression in $\varepsilon(0)$ is not due to local interaction of water with the solute but rather to the overall disruption of the H-bond network. 

The real H-bond network is not perfectly tetrahedral and contains loops and cooperative H-bonding effects. Bulk water is populated by many different types of H-bonded structures with varying lifetimes. The concept of polar nanoregions (PNRs) may be useful towards understanding this situation. Polar nanoregions are regions of dipole correlation on the nanometer scale which relax more or less independently of each other.\cite{Samara:R367} Polar nanoregions have been well characterized in several relaxor ferroelectrics, where they are found to range in size from 1 - 100 nm.\cite{Burton:91} 

The average lifetime of PNRs clearly would be quantified by the Debye relaxation time $\tau_D$ as it is computed in computer simulation. However, the relaxation time of some special structures may be much longer than others. Perhaps the most striking evidence for long lived (and long ranged) dipole correlations in water comes from the analysis of the ``site-dipole field" first introduced by Higo, et. al. in a study of SPC/E.\cite{Higo:5961,Dickey:051601,Higo:395,Higo:193,Takano:14} Additionally, there are tantalizing experimental hints of very slow relaxations in the bulk.\cite{Shelton:020201,shelton:084502,shelton:044503,shelton:9374,J10-1,J10-2}

\begin{table*}
    \begin{tabular}{c c c c c c c c}
model   & $\mu$ (D) & $Q_T$ (D$\Ang$) & $\varepsilon(0)$         & $\tau_D$ (ps)             & -$d\varepsilon / d T$ at 298K (K$^{-1}$) & $g_K$ & $G_K$\\
\hline
SPC        & 2.274 & 1.969 & 65.6(2)\cite{Fennell:6936}        & 8\cite{hochtl:4927} & 0.09(1)\cite{Fennell:6936}          & 2.48 & 3.70 \\
SPC/E      & 2.351 & 2.038 & 71.8(1),71.1(1)\cite{Fennell:6936}& 12\cite{hochtl:4927}   & 0.09(1)\cite{Fennell:6936}       & 2.52 & 3.76 \\
SPC/fw     & 2.390 & 2.017 & 78.1(2)\cite{raabe:234501}        & 10\cite{wu:024503}       &                                & 2.68 & 4.00 \\
TIP3P      & 2.347 & 1.720 & 101(2),94-100(2)\cite{Vega:19663,wu:024503} & 6\cite{wu:024503}   & 7.3(7)\cite{hochtl:4927}  & 3.46 & 5.16 \\
TIP4P      & 2.180 & 2.345 & 51 (1), 50(3)\cite{Vega:19663}    & 6 \cite{neumann:1567}        & 0.19(1)                    & 2.07 & 3.08 \\
TIP4P/2005 & 2.305 & 2.514 & 59.3(4),63(1)\cite{horn:9665}     & 13                & 0.18,0.23(1)\cite{horn:9665}          & 2.19 & 3.26 \\
TIP4P/2005f& 2.319 & 2.561 & 58.8(4),55.3\cite{gonzalez:224516}& 12                & 0.20                                  & 2.14 & 3.18 \\
TIP5P      & 2.290 & 1.565 & 81-91(5)\cite{Vega:19663,mahoney:8910,rick:6085}& 8\cite{Jia:1590} & .31(1)\cite{mahoney:8910,rick:6085}& 3.22 & 4.80 \\
TTM3F      & 2.750 & 1.986 & 94.4                              & 12                &  .46                                  & 2.45 & 3.66 \\
\hline
Exp.       &~2.95\cite{badyal:9206,gubskaya:5290}& 2.565$^*$\cite{verhoeven:3222} & 78.6\cite{fernandez:1125} & 8.3\cite{ellison:1} \cite{wu:024503}& .40 \cite{fernandez:1125}& 1.77 & 2.64\\
    \end{tabular}
    \caption{Dielectric properties for some popular empirical water models at 298/300 K. The magnitude of the quadruple moment for water is well quantified by the tetrahedral quadrupole moment $Q_T = \frac{1}{2} ( |Q_{\ff{xx}}| + |Q_{\ff{yy}}| ) $.\cite{rick:6085} $G_K$ was calculated using eqn. \ref{eqnused2} and $g_K$ was calculated using eqn. \ref{Gkgkrelation}.  Values without references are from this work. Numbers in parenthesis refer to the estimated error in the last reported digit. $^*$The experimental value for $Q_T$ is for the gas phase geometry.}
    \label{watermodels}
\end{table*}

\subsection{The importance of water model geometry}
Typically empirical models are optimized to reproduce experimental values for easily computable quantities such as the density, enthalpy of vaporization, the location of peaks in radial distribution functions and possibly one or two other variables. These optimizations have led to a considerable range of dielectric constants, as shown in Table \ref{watermodels}.  Reparameterization to fix the dielectric constant has been done for SPC/E\cite{Fennell:6936} and TIP4Q.\cite{Alejandre:19728} 

The dielectric constant is very sensitive to the equilibrium bond angle $\theta_{\ff{HOH}}^{\ff{eq}}$ and $r_{\ff{OH}}$ distance. These two parameters, along with the hydrogen charge $q_{\ff{H}}$ determine the dipole moment and quadrupole moment of the molecule for a three site model. Four and five site models contain additional geometric parameters. In general $\varepsilon (0)$ increases as $\mu^2$ and decreases with an increasing quadrupole moment $Q_T$, which disrupts dipole-dipole correlations.\cite{rick:6085} 
Increasing $r_{\ff{OH}}$ increases both the dipole moment and quadrupole moment, leading to only modest increases in $\varepsilon (0)$, since these changes act in opposite directions. Increasing $\theta_{\ff{HOH}}^{\ff{eq}}$ decreases the dipole moment and decreases the quadrupole moment, both of which act in the same direction to decrease $\varepsilon (0)$.  Increasing $\theta_{\ff{HOH}}^{\ff{eq}}$ also reduces the degree to which the H-bonded shells contribute to $\varepsilon(0)$  (see eqn. \ref{HbondShells}), which further decreases the dielectric constant. Overall, the differences in dielectric constant between rigid models can be largely accounted for by differences in $\theta_{\ff{HOH}}$ and $q_{\ff{H}}$.\cite{hochtl:4927} It is important to bear in mind that even small changes in $\theta_{\ff{HOH}}^{\ff{eq}}$ and $q_{\ff{H}}$ can have a larger effect on $\varepsilon(0)$ than the introduction of flexibility or polarizability to a model.

\section{Simulation details}
\subsection{Molecular dynamics}
To determine the effect of flexibility we choose to compare the TIP4P/2005 model of Abascal \& Vega\cite{abascal:234505} and the TIP4P/2005f model of Gonzalez \& Abascal.\cite{gonzalez:224516}  Although its value for $\varepsilon(0)$ is less accurate than other more popular empirical models (like SPC/E or TIP4P) TIP4P/2005 was recently scored as best overall among five popular rigid models.\cite{Vega:19663} In particular, it is better at reproducing the liquid structure, density-temperature curve and phase diagram. Although the value of $\varepsilon(0)$ of TIP4P/2005 is not as good as other models, it more accurately describes the variation of the dielectric constant with temperature (discussed below). 

Our TIP4P/2005 simulations were performed with the GROMACS molecular dynamics package (versions 3.3.3 and 4.5.5).\cite{Hess:435} All of our GROMACS runs used a Nos\'{e}-Hoover thermostat with $\tau = 1$ ps or $\tau = .1$ ps. For rigid simulations we used a timestep of 2 fs and for flexible simulations we used a timestep of .5 fs. The GROMACS simulations with 512 molecules (used for all dielectric constant calculations) employed a Coulomb cutoff of 1.2 nm and a shifted VdW cutoff of 1.1 nm. For the long range part of the Coulomb interaction particle mesh Ewald (PME) was employed.

For a polarizable model we choose the TTM3F model of Fanourgakis \& Xantheas.\cite{fanourgakis:074506} It is a four site model, so it has a similar geometry to TIP4P/2005. The model contains one polarization dipole per molecule located on the m-site. It also contains fluctuating charges, which are determined using a potential energy surface and dipole moment surface derived from ab-initio simulation. This fluctuation of charge is also a polarization effect, however we measured the charge fluctuations to be small (only $\pm 2 \%$ at 300 K). We determined that the contribution to the dipole fluctuation from charge fluctuation is about 4.3 times smaller than the contribution from the polarization dipole at 300 K. Our TTM3F runs used a Nos\'{e}-Hoover thermostat with $\tau = .1$ ps, a timestep of .5 fs and Coulomb and VdW cutoffs of .7 nm. The VdW cutoff was switched off using the ``GROMACS switch"\cite{GROMACSmaual} and long range VdW corrections to the energy were applied. Ewald summation was used, where the smeared dipoles and charges are considered as point dipoles and point charges. The polarization dipole was calculated using a convergence tolerance of $10^{-5}$ D per molecule. A fourth order predictor was used to provide the first guess for each iteration, reducing the number of required iterations per timestep from 15 to 2 - 3.

We ran all of our simulations in the NVT ensemble. We decided not to use a barostat largely for simplicity but also to prevent the possibility of the barostat interfering with the dynamics of the system. The NVT ensemble also allows us to analyse the effects from changes in density and effects from changes in temperature separately. 

\subsection{Calculation of dynamical quantities}
The frequency dependent dielectric constant was calculated from the dipole autocorrelation function $\Phi (t)$ using the linear response equation:
\begin{equation}\label{EpsOmegaEqn}
       \varepsilon (\omega) -  \varepsilon_\infty = (\varepsilon(0) -  \varepsilon_\infty)\mathcal{L}[-\dot{\Phi}]
\end{equation}
\begin{equation}
     \Phi (t) = \frac{\langle {\bf M}(0)\cdot{\bf M}(t)\rangle}{\langle M^2 \rangle}
\end{equation}

Here $\mathcal{L}[]$ is the ``Fourier-Laplace" (one-sided Fourier) transform:
\begin{equation}
    \mathcal{L}[f(t)] = \int_0^{\infty} dt e^{-i\omega t} f(t)
\end{equation}

The Debye relaxation time $\tau_D$ and the single molecule relaxation time $\tau_s$ were calculated by fitting an exponential to the total box and single molecule dipole autocorrelation functions, which are denoted by $\Phi (t)$ and $\phi (t)$.


The short time parts (0 - .5 ps) of $\Phi (t)$ and $\phi(t)$ exhibit a rapid decrease and oscillatory behavior due to rapid librational and vibrational motions. Sometimes this part is accounted for by fitting with two exponentials, the shorter relaxation time $\tau_2$ being called the ``second Debye relaxation". In our case we choose to simply ignore the short time behavior of $\phi (t)$ and did our fits starting at around 2 ps and going out a few ps until the correlation function was no longer converged. The fitting function was :
\begin{equation}
    f(t) = A e^{ t/ \tau}
\end{equation}
with $A$ and $\tau$ as the free parameters.

Time correlation functions of dynamical quantities are known to converge very slowly.\cite{allen89} For this reason it is essential to fit an exponential to the $\Phi(t)$ obtained from the simulation in order to properly calculate the long time part when computing $\varepsilon (\omega)$. To prevent artifacts in $\varepsilon (\omega)$ due to poor joining of the data and fit we used a cubic spline with a length of $\approx$ 1 ps. Even with a spline we found that the joining of the fit introduces noise in $\varepsilon (\omega)$ in the range $10^{13}$ to $10^{14}$ Hz. This noise can be reduced by increasing or decreasing the length of the smoothing spline but is hard to eliminate completely. Similar noise appears in the $\varepsilon(\omega)$ plots of van der Spoel, et al, who employed a linear interpolation function.\cite{spoel:10220}

\subsection{Convergence tests}
\subsubsection{Convergence of $\varepsilon(0)$}
It is well known that long simulations are required to ensure the proper convergence of $\varepsilon(0)$ in water. A comparison of five 50 ns runs shows that at least 20 ns are necessary for $1 \%$ convergence in SPC/E.\cite{Fennell:6936} Many older studies reporting $\varepsilon(0)$ did not allow enough time for adequate convergence (ie. to within 10\%)  (this is clearly seen in 1998 summary of literature values by van der Spoel, et. al.\cite{spoel:10220}) It is interesting to note that molecular dynamics simulations of acetonitrile, another polar liquid, show convergence to within $\pm 5$\% in only .4 ns.\cite{mountain:3921} It appears that the presence of hydrogen bonding slows down dipolar fluctuations and leads to longer convergence times. This is confirmed by the fact that the time required for convergence varies dramatically with temperature from 1 - 2 ns at 400 K to 100+ ns at 220 K. 

\subsubsection{Test for artifacts from thermostating}
\begin{table}
\begin{tabular}{c c c c c }
    Thermostat & $\tau$ (ps)  &Length (ns)& P (bar) & $\varepsilon(0)$ \\
    \hline
    Nos\'{e} -Hoover & .01 & 9       & 1264  &  52.5$\pm .5$      \\
    Nos\'{e} -Hoover & 1   & 10        & 1260  &  53.1$\pm .5$      \\
    Nos\'{e} -Hoover & 100 & 9       & 1265  &  53.6$\pm .5$      \\
    Berendsen        & .01 & 9         & 1261  &  54.0$\pm .6$     \\
    Berendsen        & 1   & 9       & 1265  &  53.8$\pm .6$     \\
    Berendsen        & 100 & 10        & 1367  &  53.6$\pm .4$     \\
    \end{tabular}
    \caption{Test thermostating runs at 300 K performed with 512 TIP4P.}
    \label{ThermostatRuns}
\end{table}

Previously it has been reported that changing from a Berendsen to a Nos\'{e}-Hoover thermostat resulted in an increase in $\varepsilon(0)$ of 5\%.\cite{Gereben:80} To see if thermostating has any effect on $\varepsilon(0)$ and $\varepsilon(\omega)$ a series of simulations were run at 300 K with 512 TIP4P molecules using Berendson and Nos\'{e}-Hoover thermostats with time constants of $\tau =$ .01, 1, and 100 ps. It was observed that all of the simulations maintained their temperatures properly and yielded $\varepsilon(0)$ which were equivalent within their errors (table \ref{ThermostatRuns}). No systematic dependence of $\varepsilon(0)$ on $\tau$ was discernible, nor was there any discernible difference between the Berendson \& Nos\'{e}-Hoover techniques. The previously reported discrepancy of 5 \% is likely attributable to improper convergence as their simulations were only 8 ns.\cite{Gereben:80}  When comparing $\varepsilon(\omega)$ for these simulations no noticeable differences were observed even with $\tau = .01$ ps.

Even though thermostating had no effect on $\varepsilon(0)$ or $\varepsilon(\omega)$ it was noticed that the presence of a thermostat did increase the time required for proper convergence compared to an NVE simulation. This is not surprising, especially for the Berendson thermostat which periodically rescales the velocities of molecules, interrupting cooperative fluctuations in $\bf{M}$.


\subsubsection{Test for finite size artifacts}

Whenever one does a computer simulation one should always consider the possibility of finite size effects, especially when using periodic boundary conditions to simulate a non-periodic system. For a system of dipoles on a cubic lattice with PBC and Ewald summation it has been shown that $\varepsilon(0)$ approaches the proper thermodynamic limit from below as $N^{-2/3}$.\cite{Morrow:187} To see if this is the case in water we ran a series of 20 ns TIP4P simulations at 300 K with 16, 64, 256, 512 and 1000 molecules (see supplementary material\cite{Note1}). There was no difference in $\varepsilon(0)$ between 512 and 1000 molecules, suggesting 512 is adequate. The convergence does not follow the $N^{-2/3}$ law, but the system appeared to be approaching the thermodynamic limit from below as expected.  

\section{Results for $\varepsilon(0)$}
\begin{figure}
 \centering
 \includegraphics[width=8.5cm]{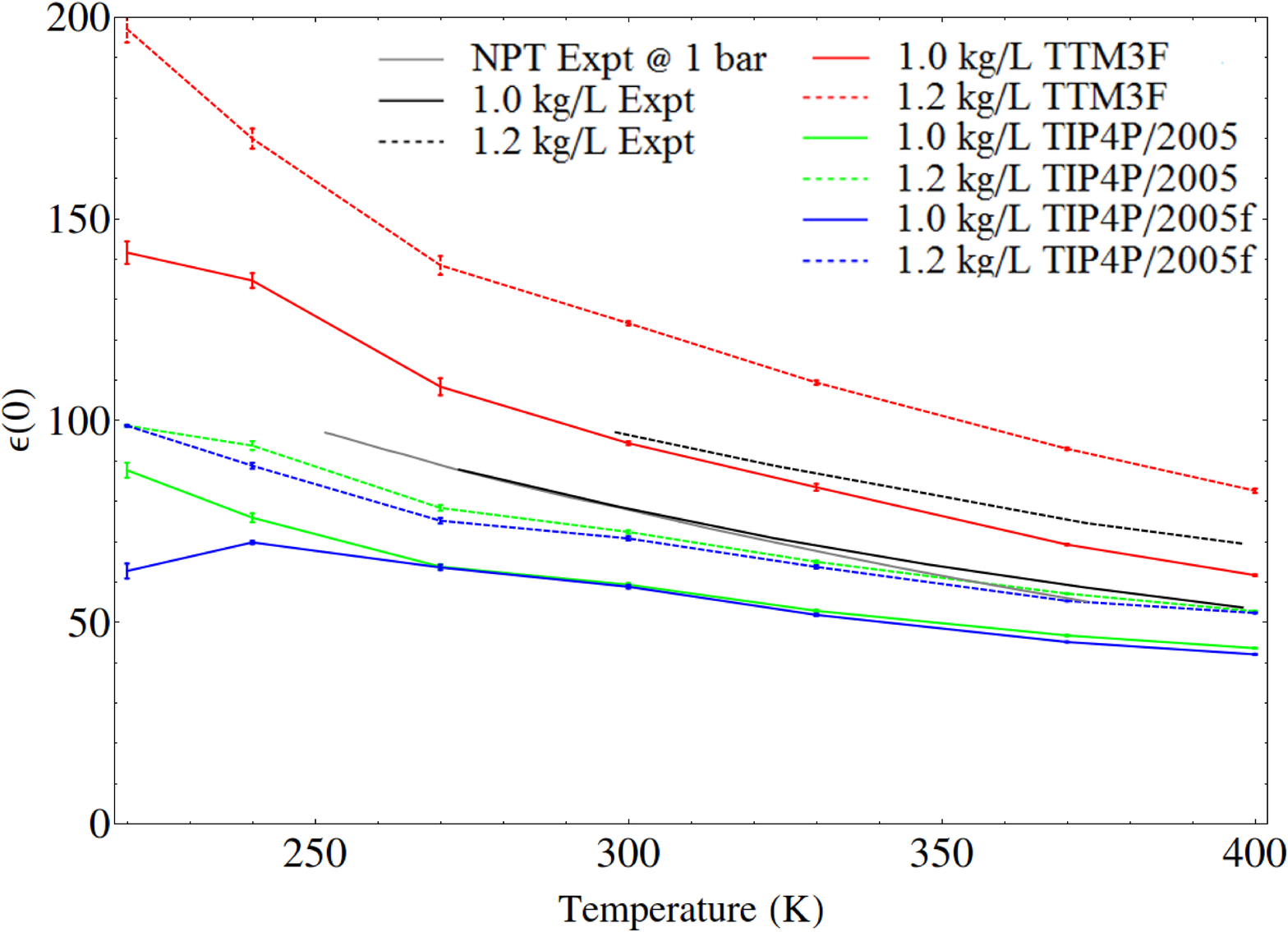}
  \caption{Dielectric constants for TIP4P/2005, TIP4P/2005f and TTM3F at 1 kg/L and 1.2 kg/L. The experimental values along the 1.0 kg/L isochore were taken by interpolating the tables given by Uematsu and Frank.\cite{uematsu:1291} The experimental values at 1.2 kg/L were obtained by extrapolating the same tables to higher pressure.}
  \label{eps0}
\end{figure}

Figure \ref{eps0} shows the dielectric constants of the three models. The experimental $\varepsilon(0)$ values along the 1.00 kg/L and 1.20 kg/L isochores are taken from the $\varepsilon(0)$ vs. pressure tables developed by Uematsu and Frank.\cite{uematsu:1291} The dependence of $\varepsilon(0)$ on pressure is very close to linear, so a linear extrapolation of the Uematsu \& Frank data was used to estimate $\varepsilon (0)$ at 1.2 kg/L. The pressure required to achieve 1.0 kg/L or 1.2 kg/L at different temperatures were taken from the ASME Steam Tables based on the IAPWS-1997 formulation,\cite{parry2006asme} which are freely accessible at wolframalpha.com. We also plotted experimental data taken along the 1 bar isobar.\cite{fernandez:1125,bertolini:3285}

At all state points the dielectric constant of TIP4P/2005 is nearly equal to that of TI4P/2005f. This lack of change should be contrasted with the changes in $\varepsilon (0)$ observed in flexible versions the SPC model. The flexible model of Wu, Tepper \& Wolf (SPC/Fw) yields a dielectric constant which is 23\% larger than SPC at STP,\cite{raabe:234501} and the flexible model of Dang \& Pettit (SPC/Fd) yields a dielectric constant which is 54\% larger.\cite{wu:024503}

In developing TIP4P/2005f, the flexibility was added in a careful manner to ensure that the geometry of TIP4P/2005 was well preserved. The percent differences in the liquid HOH angle and $r_{\ff{OH}}$ distance are only .26 \% and 1 \%.\cite{gonzalez:224516} The only other change they made was to make the Leonard-Jones $\sigma$ parameter in TIP4P/2005f a little bit (.002\%) smaller. In the SPC/Fw model of Wu et al. the flexibility was parametrized specifically to reproduce the experimental $\varepsilon (0)$ and diffusion constant $D_s$. As a result of this SPC/Fw has a smaller liquid phase $\theta_{\ff{HOH}}^{\ff{eq}}$ (107.7$^{\circ}$ vs. 109.47$^{\circ}$) and a longer $r_{\ff{OH}}$, changes of 1.6\% and 4 \%. This resulted in SPC/Fw having a larger average dipole moment ( $2.39 D$ vs. $2.275 D$ - an increase of 5 \%). The same is true in SPC/Fd, but to an even greater extent, yielding a dipole of $2.47 D$. By contrast the average dipole of TIP4P/2005f is only slightly larger than that of TIP4P/2005 ($2.319 D$ vs. $2.305 D$ - an increase of .6 \%). Another difference is that the HOH bending potential in both SPC/Fw and SPC/Fd allow greater flexibility, since the coefficient $K_\theta$ is 14\% smaller in both models.  

TTM3F has a larger dielectric constant than TIP4P/2005, despite having a slightly larger $\theta_{\ff{HOH}}$ angle (105.13$^{\circ}$ vs 104.52$^{\circ}$), which by itself would decrease the dielectric constant by a few percent.\cite{wu:024503} The increase is clearly due to a larger overall dipole moment and greater dipole-dipole correlation (discussed below). 


\begin{table}
    \begin{tabular}{c c c c c c c}
           & \multicolumn{6}{c}{Temperature (K)} \\
           & 240 & 270  & 300  & 330  & 370 & 400  \\
    \hline
TIP4P/2005 & 23  & 23   & 22   & 23   & 22  & 21    \\
TIP4P/2005f& 27  & 18   & 20   & 23   & 23  & 25    \\
     TTM3F & 31  & 28   & 31   & 31   & 34  & 35    \\
   Expt    & -   & -    & 23   & 26   & 27  & 30
    \end{tabular}
    \caption{Percentage increase in dielectric constant going from 1 kg/ L to 1.2 kg / L.}
    \label{Eps0PercentChange}
\end{table}

Increasing the density increases $\varepsilon (0)$ as can clearly be seen from equation \ref{eqnused2}. Table \ref{Eps0PercentChange} shows the percentage increase in $\varepsilon(0)$ for the three models when the density is increased to 1.20 kg /L. For both rigid and flexible TIP4P/2005 the increase is around 22\% at nearly all temperatures.  From equation \ref{eqnused2} one sees that this linear increase with density is consistent with $G_K$ not increasing with density. With TTM3F, the increase is significantly larger than 20\%, indicating that $G_K$ increases with density. Although TTM3F overestimates this increase when compared to experiment, it captures the temperature dependence of the increase correctly.

Table \ref{mudata} shows the average dipole moments of TIP4P/2005f and TTM3F at the two densities. The increase in the dipole moment of TTM3F with density is almost completely due to an increase in the polarization dipole.

\begin{table}
    \begin{tabular}{c c c c}
 density (kg/L)            &      1.00       &    1.20         & \% increase \\
\hline
TIP4P/2005f                & 2.319$\pm 0.14$ & 2.323$\pm 0.14$ & .1 \\
TTM3F total dipole         & 2.750$\pm 0.19$ & 2.785$\pm 0.24$ & 1.2\\
TTM3F polarization dipole  & 0.827$\pm 0.16$ & 0.857$\pm 0.16$ & 3.6\\
TTM3F geometric dipole     & 1.922           &  1.927          & .2 \\
    \end{tabular}
    \caption{Average dipole moments and their standard deviations for TIP4P/2005f and TTM3F. }
    \label{mudata}
\end{table}

\subsection{Temperature derivative of $\varepsilon(0)$}
\begin{figure}
  \includegraphics[width=8.5cm]{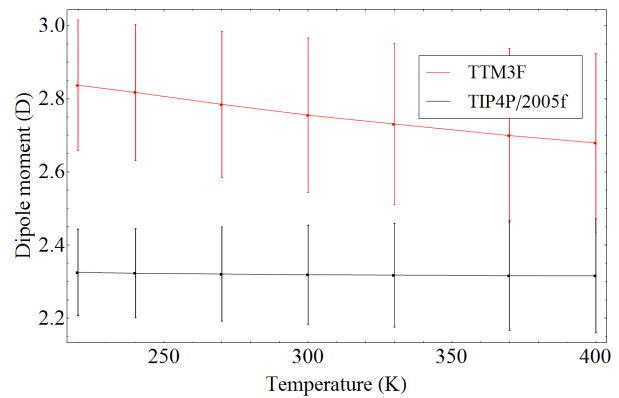}
  \caption{Average dipole moments for TTM3F and TIP4P/2005f vs. temperature at a fixed density of 1 kg/L. The error bars show the standard deviations of the dipole moment distributions. The results show that the addition of polarization leads to a temperature dependent dipole moment, even when the density is fixed.}
   \label{MuVsTemp}
\end{figure}
\begin{figure}
 \centering
 \includegraphics[width=8.5cm]{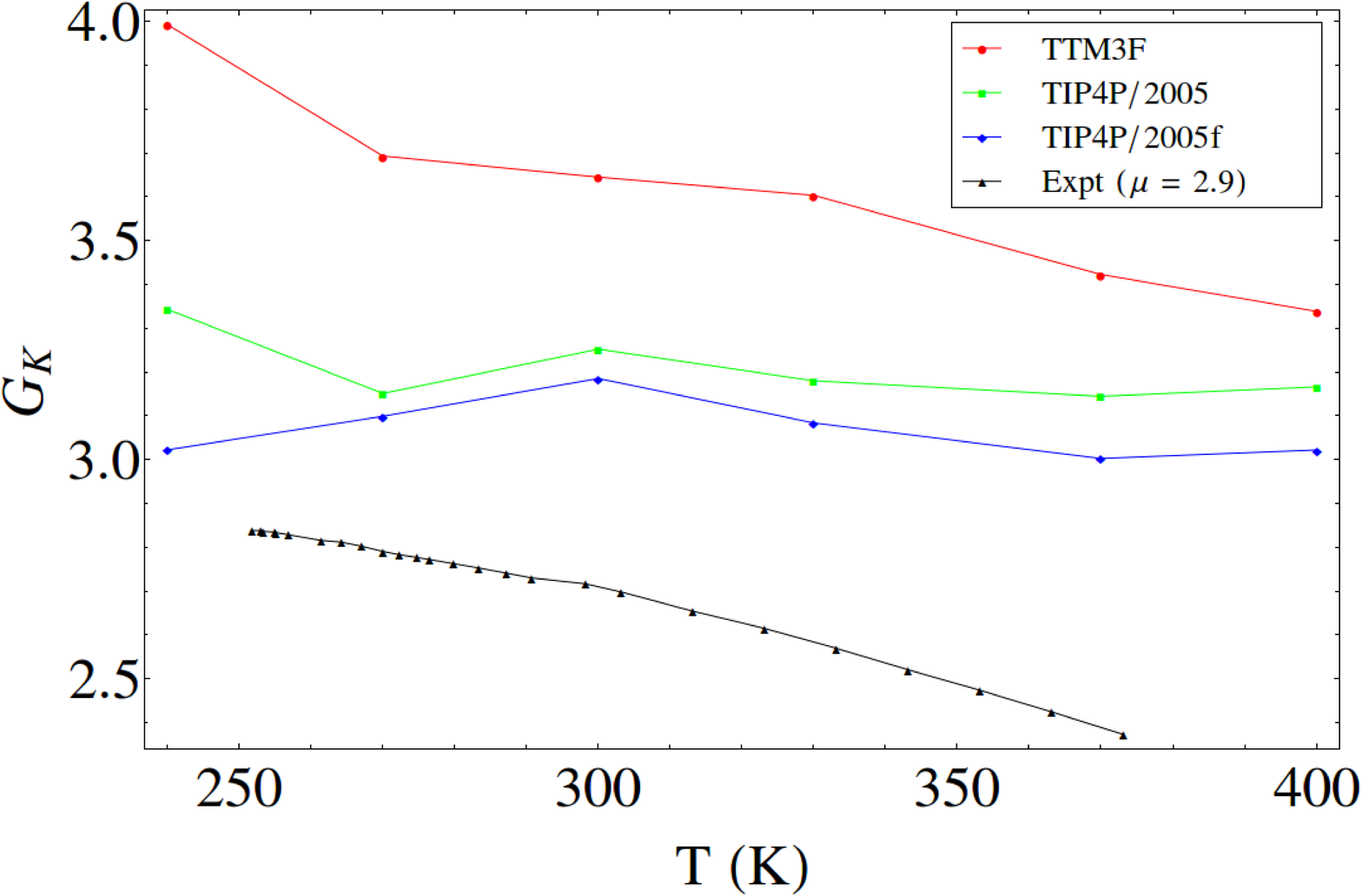}
  \caption{$G_K(r)$ for the models at different temperatures, calculated using $\varepsilon(0)(T)$ and $\mu(T)$. The experimental data was calculated using experimental $\varepsilon(0)$\cite{fernandez:1125,bertolini:3285} using eqn. \ref{eqnused2} and $\mu = 2.9$.}
  \label{GKvsT}
\end{figure}

The temperature derivative of $\varepsilon(0)$ is an important quantity which has been largely neglected in studies of water models. The temperature derivative is directly proportional to the change in entropy of the liquid under the application of an electric field.\cite{F49}\cite{surfaceforces3} Thus an accurate value of $d \varepsilon(0) / dT $ is important for capturing the change in the entropy (ordering) of the liquid around ions and predicting the solvation free energy of charged species.\cite{surfaceforces3} For this reason $d \varepsilon(0) / dT $ at 298/300K is compared for some popular water models in table \ref{watermodels}. Interestingly, SPC/E greatly underestimates $d \varepsilon(0) / dT $ while TIP3P overestimates it. SPC/E and TIP3P are the two most popular explicit water models in the biophysics community.\cite{Note2} Of the water models listed, TTM3F most accurately captures the slope at 300 K.

It is also useful to look at the temperature dependence of $G_K$ when comparing the models (see figure \ref{GKvsT}). All three models overestimate the degree of correlation but TTM3F yields the correct monotonic decrease in $G_K$ with increasing temperature, while TIP4P/2005 and TIP4P/2005f show an unphysical increase in $G_K$ with temperature between 240 and 300 K and then little change at higher temperatures. 

TTM3F exhibits temperature dependence of $\mu$ even at fixed density, as shown in figure \ref{MuVsTemp}. This is likely the distinguishing factor which allows TTM3F to have a better temperature derivative compared to the other models.

\section{Results for $\varepsilon(\omega)$}
\begin{figure}
  \includegraphics[width=8.5cm]{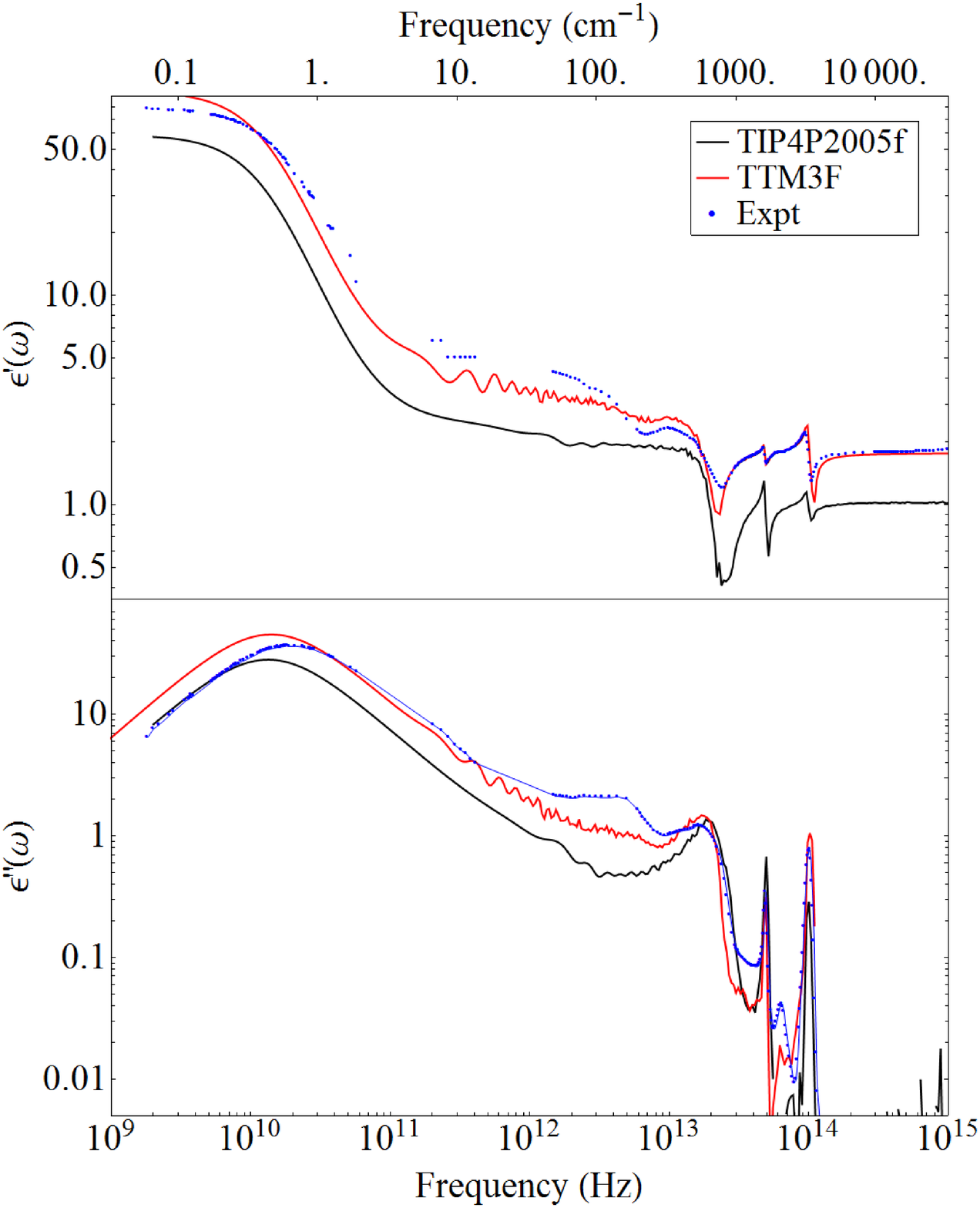}
  \caption{Real part (top) and imaginary part (bottom) of the dielectric spectra at 300 K. The region between 10 to 100 cm$^{-1}$ is plagued by noise from the fitting process.}
  \label{EpsOmega}
\end{figure}
\begin{figure*}
  \includegraphics[width=18cm]{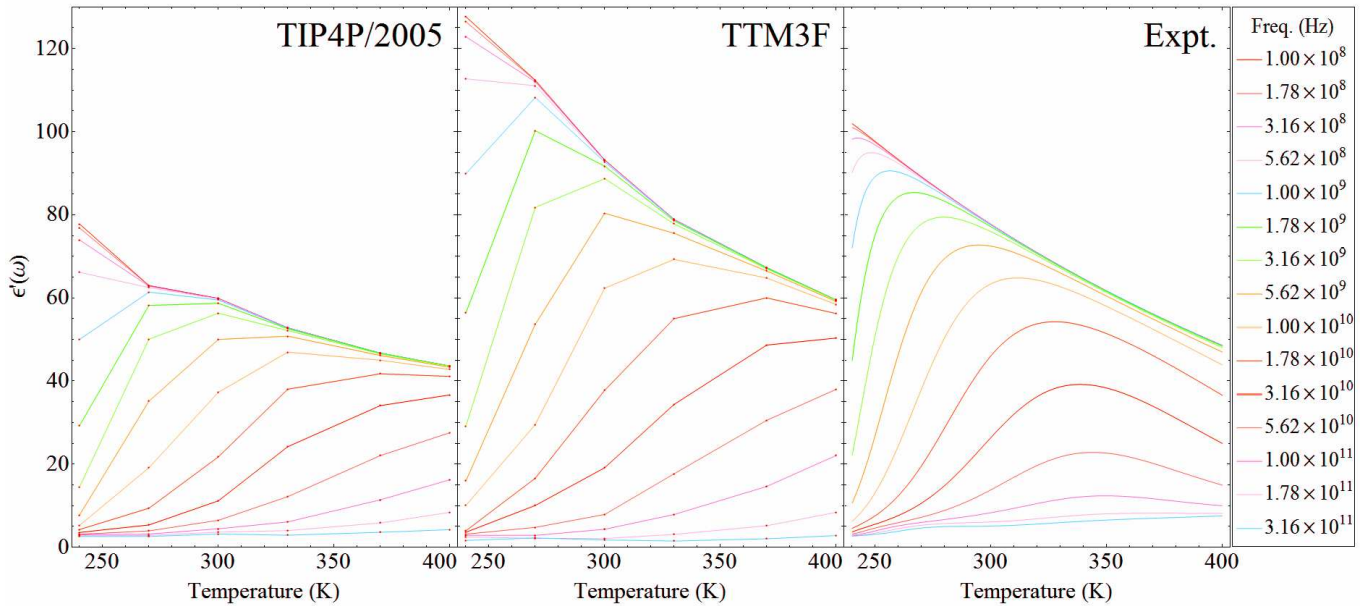}
  \caption{The temperature dependence of $\varepsilon'(\omega)$ at different frequencies. The experimental data is a two-Debye fit function $\varepsilon'(\omega,T)$ derived from experimental data by Meissner and Wentz.\cite{Meissner:1836} It was shown to very accurately reproduce experimental measurements between 273 and 373 K.}
    \label{TempDependence}
\end{figure*}

%
%

Figure ~\ref{EpsOmega} shows the real and imaginary dielectric functions. The experimental data between 
50 - 33,333 cm$^{-1}$ (1.5 $\times$ 10$^{11}$ -  10$^{15}$ Hz) 
was taken from index of fraction data using the relation $\varepsilon (\omega) = n^2(\omega)$.\cite{Hale:73}

Of particular interest is the the feature centered at 180-200 cm$^{-1}$ which is most clearly present in $\varepsilon''(\omega)$. Neumann noted that this feature is absent in the dielectric spectra of TIP4P and proposed that it must be due to polarization effects.\cite{neumann:1567} Raman and FIR spectra of water also show a band between 170-190 cm$^{-1}$.\cite{Fukasawa:197802,Nielson:3,Madden:502,Hasted:622}

The exact nature of the 180 cm$^{-1}$ Raman band has been the subject of some controversy.\cite{Nielson:3} The prevailing view is that it is due to the stretching vibrations of nearly-linear hydrogen bonds, but others have interpreted it as being due to cage vibrations or more exotic hydrogen bond network relaxations.\cite{Nielson:3}


If the feature at 180-200 cm$^{-1}$ is indeed due to the stretching of hydrogen bonds, then it will only appear in $\varepsilon(\omega)$ if polarization is included, as the geometric dipoles of two H-bonded molecules do not change during H-bond stretching. Indeed, the TTM3F spectrum shows a shallow peak in this region, while the flexible TIP4P/2005 shows nothing. The fact that the TTM3F peak is smaller than experiment makes sense considering that the hydrogens are not polarizable in TTM3F and the only polarization dipole is located on the m-site.  

At high frequencies we see that both TTM3F and TIP4P/2005f do a good job of reproducing the librational resonances and the bending ($v_2$) and symmetric \& antisymmetric modes ($v_1 + v_3$), with TTM3F performing noticeably better in reproducing $\varepsilon''(\omega)$.  Using the Clausius-Mossotti equation we calculated $\varepsilon_\infty$ for TTM3F to be 1.76 using the polarizability of the polarization dipole only. The molecular polarizability from flexibility was estimated by calculating the change in dipole due to bending in an electric field oriented along the HOH bisector. For TIP4P/2005f we found $\varepsilon_\infty \approx 1.04 $.

\subsection{Temperature dependence of $\varepsilon(\omega)$}
Figure \ref{TempDependence} shows the temperature dependence of the real part of the dielectric constant $\varepsilon'(\omega)$ at different frequencies. To our knowledge such plots have only been presented once before for water, on the website of M. Chaplin.\cite{Chaplin:1}

According to a review article on relaxor ferroelectrics, ``a universal signature of the relaxor state is a broad frequency-dependent peak in the real part of the temperature-dependent dielectric susceptibility".\cite{Samara:R367} The ``experimental data" here comes from a two-Debye fit function for $\varepsilon'(\omega, T)$ derived from experimental data by Meissner and Wentz.\cite{Meissner:1836,Chaplin:1} It was shown that this fitting function well reproduces the experimental data for $\varepsilon'(\omega,T)$ between 273 and 373 K. Of particular interest is the temperature dependence of the $\varepsilon'(\omega)$ peak, which is better captured by TTM3F.

\section{The dipolar relaxation time}
\begin{figure}
  \centering
  \includegraphics[width=8.5cm]{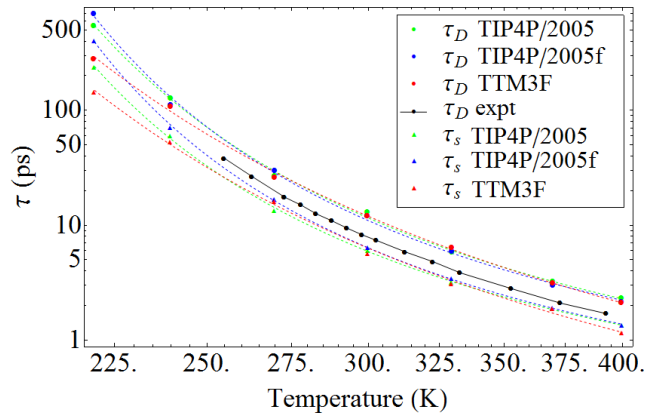}
    \caption{Relaxation times for the entire box (squares) and for a single molecule (triangles). VFT fits are shown as dashed lines. To improve the quality of these fits they were done logarithmically, as is a standard procedure for producing exponential fits. Otherwise, the least squares minimization is dominated by the lower temperature data and the higher temperature is not fit. The spread in the points at low temperature is likely due to incomplete convergence of the correlation functions due to the glassy nature of the system.}
    \label{TausComparison}
\end{figure}

The temperature dependence of both $\tau_D$ and $\tau_s$ is best described by the Vogel-Fulcher-Tammann (VFT) equation:
\begin{equation}
    \tau = \tau_{\infty} \exp \left( \frac{  D T_{\ff{VFT}} }{  T - T_{\ff{VFT}}    } \right)
\end{equation}
(See the supplementary material for a comparison of the VFT fit with other fitting functions.\cite{Note3}) This fact is very interesting because VFT relaxation is a universal feature of relaxor ferroelectrics and dipolar glasses.\cite{Pirc:020101,Bokov:4899} The underlying origin of the VFT equation is not very well understood, but most theories assume a distribution of relaxation environments within the bulk. An influential theory for the VFT equation is the Adam-Gibbs model, which assumes the existence of cooperatively rearranging regions.\cite{adam:139} The cooperatively rearranging concept is nearly identical to the polar nanoregion concept used to describe relaxor ferroelectrics. The fact that the Debye relaxation is larger than the single molecule relaxation function is a direct consequence of dipolar correlations. A model which assumes spherical relaxation clusters (analogous to PNRs) predicts $\tau_D / \tau_s = 3 G_K$.\cite{Arkhipov:127}
	A comparison of the three models studied here shows little difference in $\tau_s(T)$ or $\tau_D(T)$ between the models (figure \ref{TausComparison}). Thus the introduction of polarization does not appear to significantly effect the Debye or single molecule relaxation times.

\section{Relaxation at different length scales}
\begin{figure}
  \centering
  \includegraphics[width=8.5cm]{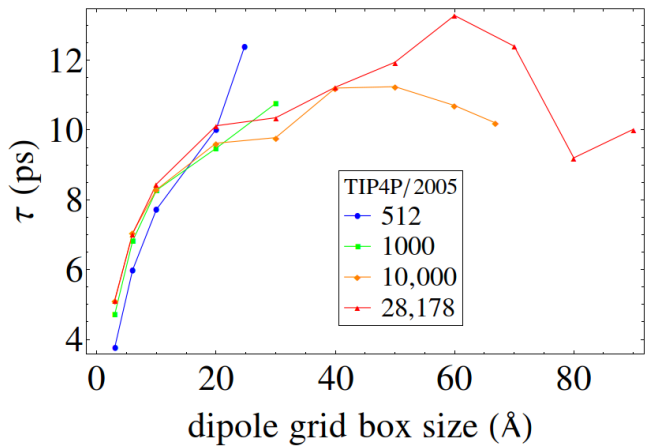}
  \caption{Values of $\tau$ for sub-boxes of different sizes, TIP4P/2005 at 300 K. The total boxes contained either 512, 1,000, 10,000 or 28,178 molecules. For the 512 molecule box the sub-boxes had sizes of $L =$ 3, 6, 10 and 24.8 $\Ang$ corresponding to boxes with approximately 1, 7, 33 and 512 molecules. Care was taken to use a consistent fitting procedure. The error was estimated to be 5\% or less.}
  \label{tau_vs_size}
\end{figure}

The relaxation times of sub-boxes of different sizes gives information about the size of the polar nanoregions responsible for the Debye relaxation time. We broke the simulation cell into boxes of different sizes and calculated the total dipole moment of each box at each timestep. The dipole correlation function is computed separately for each box and then averaged over all boxes. Figure \ref{tau_vs_size} shows the resulting dependence of of the relaxation time $\tau$ on the box size. A convergence of $\tau$ appears to be reached when $L= 40 \Ang$, however beyond this $\tau$ begins to decrease in large boxes. The reason for this decrease is unknown, but is likely due to the artifact from periodic boundary conditions (discussed below), which causes decorrelation at long distances. Averaging over non-overlapping spheres with diameter $L$ gives the same result (not shown).

\section{1D angular correlation functions}
\begin{figure}
  \includegraphics[width=8.5cm]{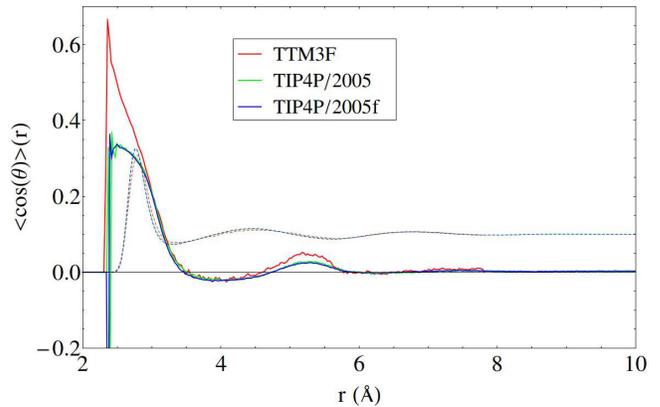}
  \caption{$\langle \cos (\theta) \rangle$ for the three models at 300 K. The O-O RDFs (rescaled by a factor of .1) are shown for comparison.}
  \label{costheta}
\end{figure}
\begin{figure}
  \includegraphics[width=8.5cm]{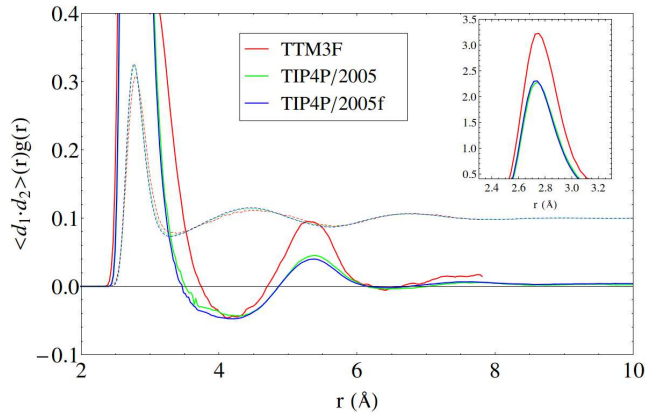}
  \caption{The dip-dip correlation function defined by equation \ref{DipDipDef}. The O-O RDFs (rescaled by a factor of .1) are shown for comparison.}
  \label{dipdipcorr}
\end{figure}
\begin{figure}
  \includegraphics[width=8.5cm]{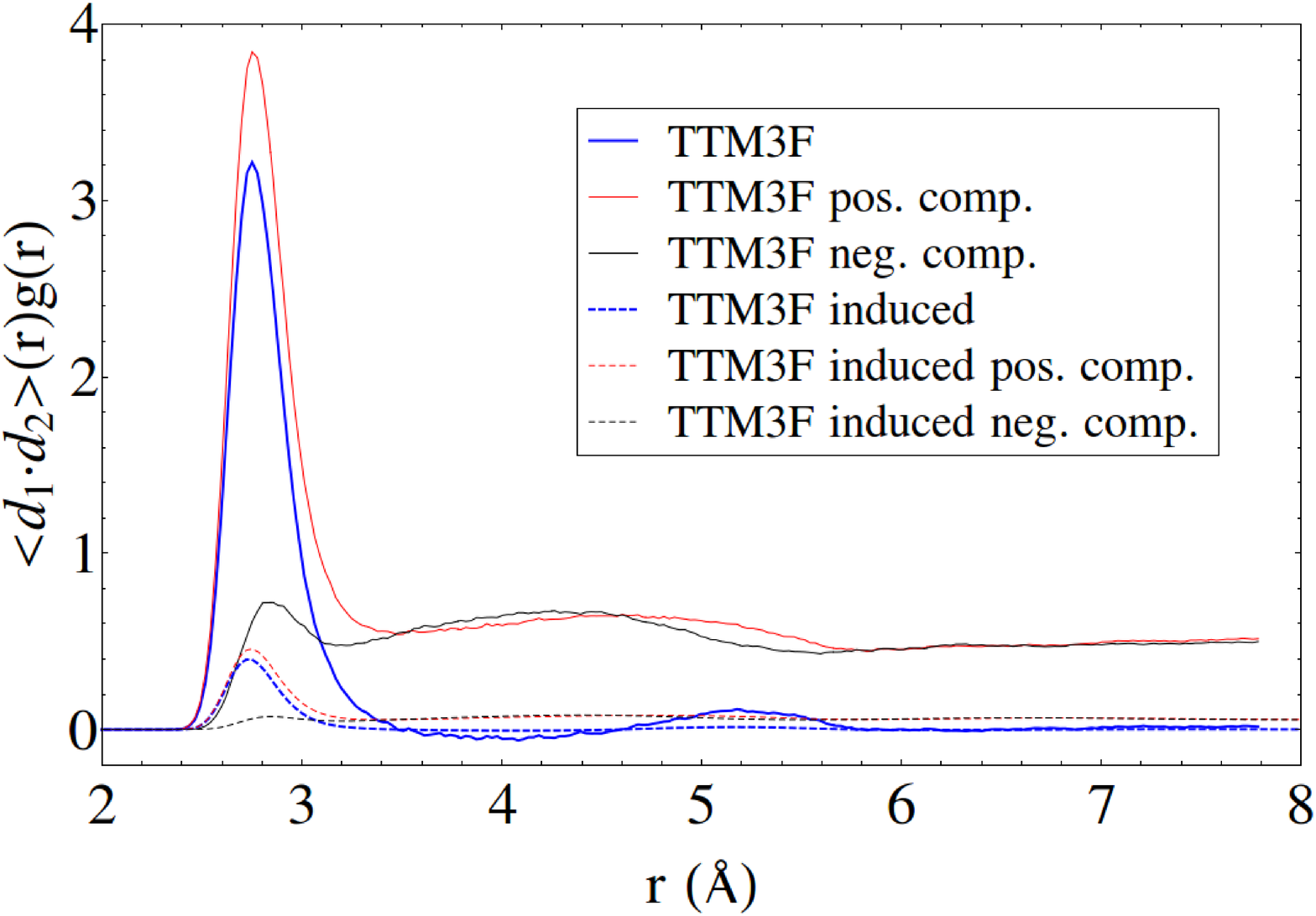}
  \caption{Positive, negative and induced components of the dip-dip correlation function for TTM3F.}
  \label{TTM3Fposneg}
\end{figure}
\begin{figure}
  \includegraphics[width=8.5cm]{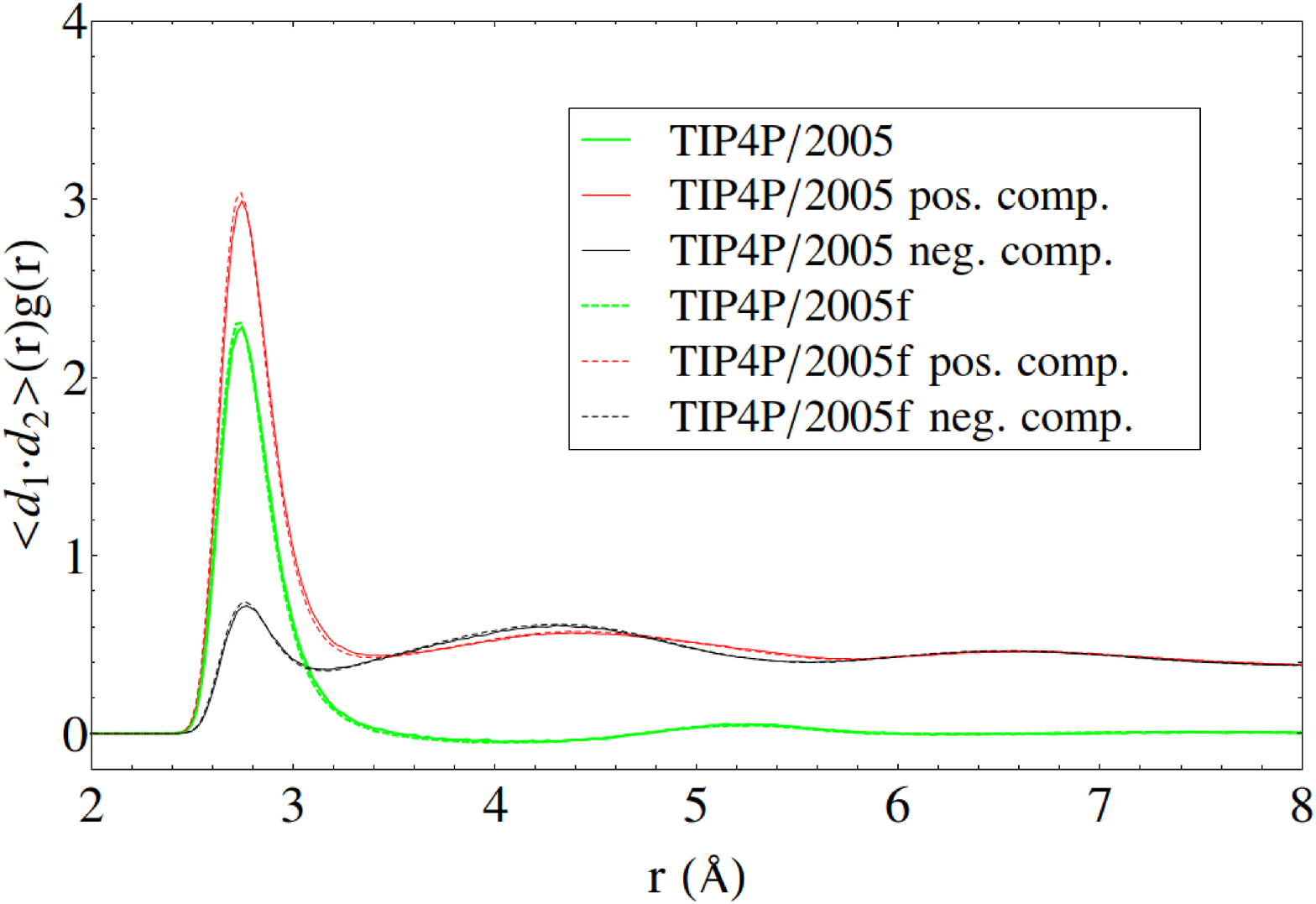}
  \caption{Positive and negative components of the dip-dip correlation function for the rigid (solid) and flexible (dashed) versions of TIP4P/2005. The rigid and flexible curves nearly overlap.}
  \label{TIP4P/2005posneg}
\end{figure}
\begin{figure}
  \includegraphics[width=8.5cm]{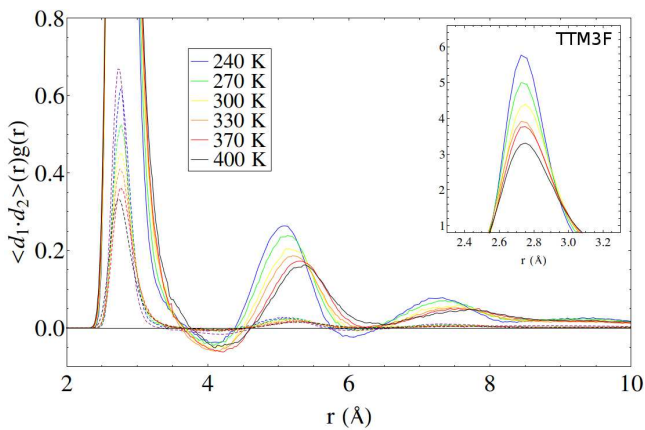}
  \caption{Dip-dip correlation function at different temperatures for TTM3F. Dashed lines show the contribution of the polarization dipoles.}
  \label{difftemp1}
\end{figure}
\begin{figure}
  \includegraphics[width=8.5cm]{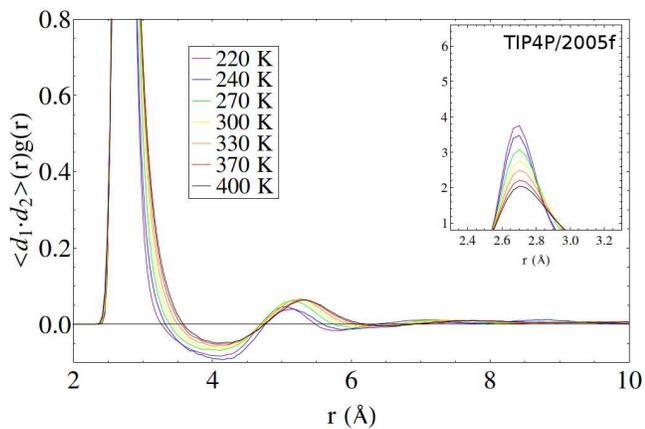}
  \caption{Dip-dip correlation function at different temperatures for TIP4P2005f.}
  \label{difftemp2}
\end{figure}
In this section we investigate two 1D correlation functions which we call the cosine function and the dip-dip correlation function. The cosine function simply gives the average cosine of the angle between the dipole moments of two molecules as a function of $r$:
\begin{equation}\label{CosDef}
    \langle \cos(\theta) \rangle (r)  = \frac{1}{N(r)} \sum'_{i,j}  \frac{ \bs{\mu}_i \cdot \bs{\mu}_j }{ |\bs{\mu}_i||\bs{\mu}_j| }  \quad r < r_{ij} <  r + \delta r
\end{equation}
The prime on the summation indicates that we do not include $i = j$. In everything that follows, angle brackets indicate an ensemble average. The cosine function is shown in figure \ref{costheta}. Oxygen-oxygen RDFs are shown for reference to emphasize that the peaks in the cosine function do not necessarily overlap with the RDF peaks, since the cosine function does not contain any information about the density of molecules. We clearly see that TTM3F has much larger correlation, especially in the first shell.  

The dip-dip correlation function is defined by:
\begin{equation}\label{DipDipDef}
    \begin{aligned}
         \phi(r) &=   \frac{1}{N_{\ff{gas}}(r)} \sum_{i,j}' \bs{\mu}_i \cdot \bs{\mu}_j    \quad r < r_{ij} <  r + \delta r\\
                 &=   \langle \bs{\mu}_1 \cdot \bs{\mu}_2 \rangle(r) g_{\ff{OO}}(r)
    \end{aligned}
\end{equation}

Here $N_{\ff{gas}}(r)$ is the number of molecules that would be found in a shell of thickness $\delta r$ at radius $r$ for a homogeneous ``gas" ($N_{\ff{gas}} (r) = 4/3\pi [ (r+\delta r)^3 - r^3] N/V$). The dip-dip correlation function for the different models at 300 K is shown in figure \ref{dipdipcorr}. Figures \ref{TTM3Fposneg} and \ref{TIP4P/2005posneg} show different contributions to the dip-dip correlation function, including the positive and negative components and (for TTM3F) the contribution of the induced dipoles.

From inspection of the first peak we see that the first H-bonded shell contributes a large positive component as expected. The region of the second H-bonded shell (4 - 5 $\Ang$) contains both positive and negative contributions. In such plots it is difficult to distinguish the contributions from H-bonded shells and non H-bonded shells, since they overlap considerably. It appears that the first interstitial shell contributes significantly to the minima at 4 $\Ang$.

Figures \ref{difftemp1} and \ref{difftemp2} compare the dip-dip correlation functions at different temperatures for TTM3F and TIP4P/2005f. TTM3F exhibits more dramatic temperature dependence and a more clearly pronounced 3rd peak. By contrast, the third peak is almost non-existent in TIP4P/2005f. The expected temperature dependence of the dipole correlation is in the expected direction in TMM3F -- ie. enhanced correlation at lower temperatures. This behaviour is not captured by either TIP4P/2005 or TIP4P/2005f, which shows less correlation in the 2nd shell at lower temperatures. 

The polarization dipoles in TTM3F contribute mainly in the first shell, where they have a large positive component. Beyond that the polarization dipoles contribute nearly equal positive and negative components which nearly cancel out. The result is a small positive contribution to the second peak and almost zero contribution to the third peak.

\section{Distance dependent Kirkwood function}
\begin{figure}
  \centering
  \includegraphics[width=8.5cm]{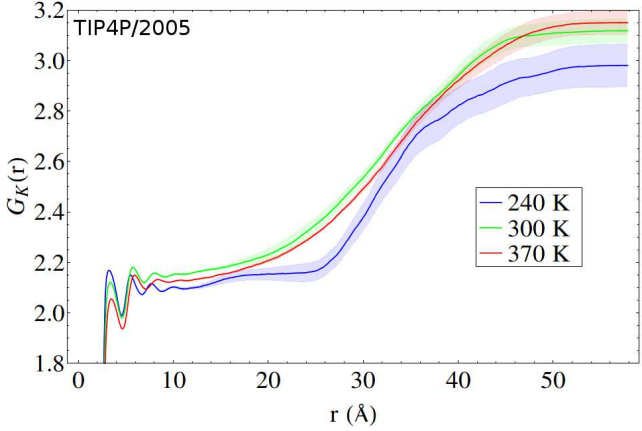}
  \caption{$G_K(r)$ function at three different temperatures for 10,000 TIP4P/2005 ($L = 66.9 \Ang$). The shaded regions show the estimated error. The dipolar ordering becomes longer ranged at lower temperatures, but also decreases in magnitude, leading to the wrong temperature dependence in $G_K$. }
   \label{LargeGkr}
\end{figure}
\begin{figure}
  \includegraphics[width=8.5cm]{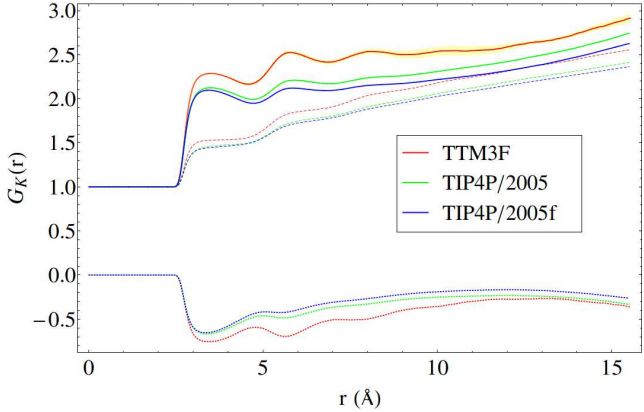}
  \caption{$G_K(r)$ functions for the three models showing the axial (dashed) and equatorial (dotted) components. Estimated errors are shown in yellow for TTM3F (the other errors were negligible). All $G_K(r)$ data beyond $\approx 9 \Ang$ is unphysical, as is discussed further in the supplementary material.}
  \label{gkrComparison}
\end{figure}
\begin{figure*}
  \centering
  \includegraphics[width=18cm]{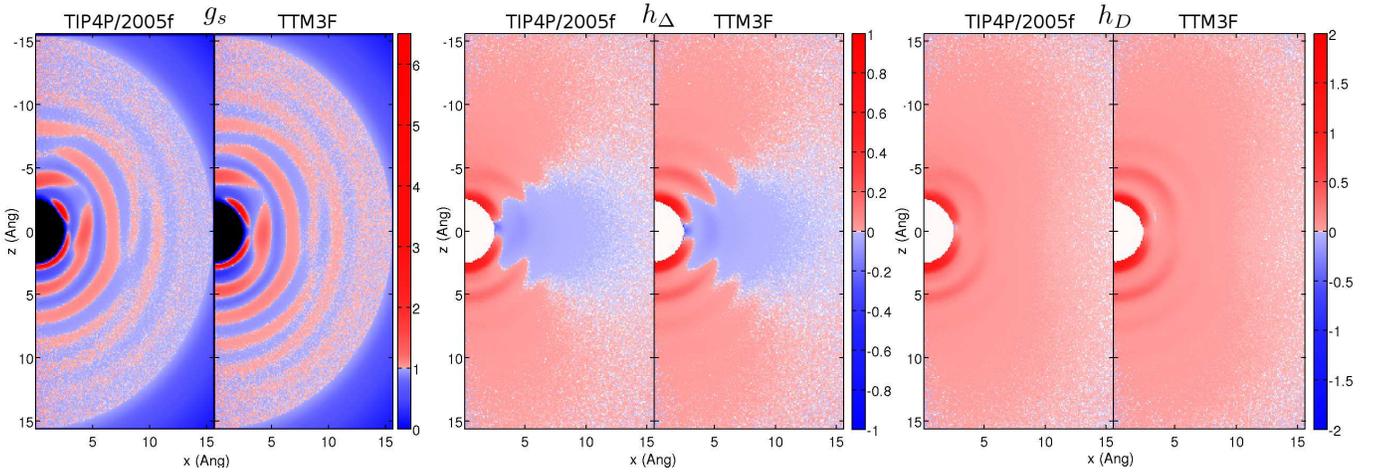}
  \caption{Comparison of 1000 TIP4P/2005f (left panels) with 1000 TTM3F (right panels). The three 2DRDFs correspond to the 2D O-O RDF (left), the 2D cosine function (middle) and the 2D dipole-dipole energy function (right). Each pixel represents a square histogram bin with $L = .1 \Ang$.}
    \label{2DRDF1}
\end{figure*}

Perhaps the most physically meaningful measure of dipole correlation is the distance dependent Kirkwood function, since it can be directly related to the dielectric constant via equation \ref{eqnused2}. For a single molecule, $G_K(r)$ is given by:
\begin{equation}
     G_K(r) = \frac{   \sum_j \bs{\mu}_1 \cdot \bs{\mu}_j }{ \langle \mu^2 \rangle}, \quad r_{1j} < r \\
\end{equation}
Averaged over $N$ molecules and all timesteps, $G_K(r)$ becomes:
\begin{equation}
     G_K(r) = \frac{  \left\langle \sum_{i,j} \bs{\mu}_i \cdot \bs{\mu}_j  \right\rangle }{ N \langle \mu^2 \rangle}, \quad r_{ij} < r
\end{equation}
The previous two dipole correlation functions become very small beyond the second shell. However, even small correlations beyond the second shell may be important as the number of molecules participating in these correlations grows as $r^2$. The Kirkwood correlation function accounts for this by reporting the total correlation of dipoles in a sphere of radius $r$ normalized only by the dipole moment of the central molecule.

Since $G_K(r)$ is more sensitive to small correlations at large distances, it is also more sensitive to artifacts arising from the use of periodic boundary conditions (PBCs) and Ewald summation.\cite{mathias:4393,mathias:10847,Spoel:1} When PBCs and Ewald summation are used $G_K(r)$ begins to artificially grow beyond a certain point which we found is usually around half the minimum image distance ($L/4$). The artifact is most clearly differentiated from the physical data in very large simulations (fig \ref{LargeGkr}). While the artifact appears large in such plots, it is accounted for in eqn. \ref{eqnused} and is locally very small.

To obtain a physically accurate $G_K(r)$ simulations of at least a few ns should be run 
in a box containing at least 5,000 molecules to cleanly separate the artifact from the data.
 Unfortunately such calculations are computationally impractical for TTM3F, so simulations of
1,000 molecules were run with lengths of 1.75 ns for TTM3F and 8 ns for TIP4P/2005 and TIP4P/2005f
(figure \ref{gkrComparison}). With 1000 molecules all $G_K(r)$ data beyond $\approx 10 \Ang$ is unphysical.
The $G_K(r)$ data clearly shows the relative contributions from different H-bonded shells to $G_K$ and therefore
 to the dielectric constant. 
Flexibility decreases $G_K(r)$ slightly in TIP4P/2005, which might be due to a weaker H-bond network.
 On the other hand, TTM3F $G_K(r)$ exhibits larger $G_K(r)$ values and displays a more pronounced contribution from the second shell.
 The third and fourth shells do not contribute to $G_K$ in any of the models but appear more pronounced in TTM3F.

Further insight can be gained by breaking $G_K(r)$ into axial and equatorial components:\cite{mathias:4393}
\begin{equation}
    G_K(r) = G_K^a(r) - G_K^e(r)
\end{equation}
If a dipole is embedded in a homogeneous dielectric continuum, the axial region is a region of positive correlation, while the equatorial is anti-correlated. The two regions are separated by a conical surface at an angle of $\theta_c = \arcsin \left( \sqrt{ \frac{2}{3} } \right) \approx 54.74 \deg$. Unlike the continuum case, the equatorial component is {\it correlated} (negative) in water out to about 1.2 nm due to the H-bond network (fig. \ref{LargeGkr}). In our simulations with 1,000 molecules the equatorial component remains negative even at large distances due to the artifact (fig. \ref{gkrComparison}). 

\section{2D angular correlation functions}

The one dimensional angular correlation functions are useful for measuring the overall correlation in each shell but do not contain any information about the structure within shells. Fully capturing the geometric correlations between molecules requires calculating the full pair correlation function $g(1,2)$ which has (for a rigid non-linear molecules) seven dimensions - a distance $r$ and three angles for each molecule (ie. Euler angles). Thus some reduction of information is necessary and many different reductions are possible. To better understand the structure we use follow the approach of Matthias \& Tavan\cite{mathias:4393} to produce 2D plots using two variables - a radial distance $r = |\bs{r}_{ij}|$ between molecules and the angle $\theta$, which is the angle between the dipole moment of molecule $i$ and $\bs{r}_{ij}$. Here $\theta = 0$ corresponds to the direction of the dipole moment (axial direction), which is called the ``z" axis. The ``x" axis lies in the plane perpendicular to the z axis (the equatorial plane). Producing this 2D plot is equivalent to doing cylindrical averages over the angle $\phi$, the equatorial angle.

Following Mathias \& Tavan we use the three ``basis functions" introduced by Wertheim:\cite{mathias:4393}
\begin{equation}\label{WertheimBasis}
    \begin{aligned}
        S &\equiv 1 \\
        \Delta &\equiv \hat{\bs{\mu}}_1 \cdot \hat{\bs{\mu}}_2 \\
        D &\equiv 3(\hat{\bs{\mu}}_1 \cdot \hat{r})(\hat{\bs{\mu}}_2 \cdot \hat{r}) - \hat{\bs{\mu}}_1 \cdot \hat{\bs{\mu}}_2
    \end{aligned}
\end{equation}
These three functions are used to make three correlation functions:
\begin{equation}
    \begin{aligned}
    g_s(r,\theta)      &\equiv \frac{V}{N^2}\left\langle  \sum_{ij} S \delta(r - r_{ij})\delta(\theta - \theta_{ij}) \right \rangle \\
    h_\Delta(r,\theta) &\equiv \left\langle  \sum_{ij} \Delta_{ij} \delta(r - r_{ij})\delta(\theta - \theta_{ij}) \right\rangle \\
    h_D (r,\theta)     &\equiv \left\langle  \sum_{ij} D_{ij}     \delta(r - r_{ij})\delta(\theta - \theta_{ij}) \right\rangle \\
    \end{aligned}
\end{equation}
The function $g_s$ is a two dimensional radial distribution function, $h_\Delta$ is a two dimensional analog of cosine function and $h_D$ gives the angular dependence of the energy of interaction (positive $h_D$ correspond to lower energies).

Figure \ref{2DRDF1} shows a comparison of the 2D correlation functions for TTM3F and TIP4P/2005f. The rigid and flexible versions of TIP4P/2005 are not compared here since they are nearly identical in appearance. Perhaps the most striking thing about these plots is their similarity -- differences in magnitude are not very visible here. Several small differences can be observed, however. The first shell in TTM3F is more spread out and thus has a smaller maxima (6.52 vs. 9.35). The TTM3F 2D cosine function exhibits slightly more structure and anti-correlation in the interstitial regions. 

In the supplementary material 2D correlation functions for 1000 SPC/E and TIP3P are also presented.\cite{Note4} In all five of the models presented the dipole correlations resemble a dielectric continuum at distances larger than 1.5 nm, confirming the findings of Mathias \& Tavan. We propose that this distance corresponds to the largest possible radius of the polar nanoregions. A sphere with $r =  1.5$ nm contains around 424 molecules. A similar maximum radius can also be deduced from the $\tau$ vs box size data or from $G_K(r)$. 

\section{Conclusions}
The results indicate that the addition of flexibility to a model, when no other reparameterizations are done, has little effect on the dielectric properties except at high frequencies. The introduction of polarization, however, does have a significant effect in several regards. Firstly, it introduces significant temperature and density dependence to the the dipole moment resulting in better values for $d\varepsilon (0) / dT $ and $d \varepsilon (0) / dV $. An accurate value for $d\varepsilon (0) / dT $ ensures that the entropy change in an electric field is described accurately, even at fixed temperature. Secondly, polarization better reproduces $\varepsilon (\omega)$, especially the 200 cm$^{-1}$ H-bond stretching feature and high-frequency features. Finally, polarization enhances dipole correlation and leads to a more physically accurate change in dipole correlation with temperature. This indicates that {\it ab initio} molecular dynamics simulations of liquid water will have larger dipole correlations. As a consequence, the analysis of local dipolar order in the form of polar nanoregions might be relevant to understanding such simulations.

\section{Acknowledgements}
This work was partially supported by DOE Award No. DE-FG02-09ER16052 (D.E) and by DOE Early Career Award No. DE-SC0003871 (M.V.F.S.). We acknowledge important discussions with Jorge I\~{n}iguez and Matthew Dawber.

%

\end{document}